\documentclass[12pt,usenatbib]{mn2e}
\usepackage{graphicx}
\usepackage{epsf}
\usepackage{color}
\usepackage{rotating}
\def\lsim{\mathrel{\lower0.6ex\hbox{$\buildrel {\textstyle <}
 \over {\scriptstyle \sim}$}}}
\def\gsim{\mathrel{\lower0.6ex\hbox{$\buildrel {\textstyle >}
 \over {\scriptstyle \sim}$}}}

\begin{document}

\title[The preferential infall of satellite galaxies]{The preferred direction of infalling satellite galaxies in the Local Group}
\author[Libeskind et al.] 
{Noam I Libeskind$^1$, Alexander Knebe$^2$, Yehuda Hoffman$^3$,  Stefan
      Gottl\"ober$^1$,  \newauthor Gustavo Yepes$^2$, Matthias Steinmetz$^1$\\
        $^1$Astrophysikalisches Institut Potsdam, An der Sternwarte 16, D-14482 Potsdam, Germany\\
  $^2$Grupo de Astrof\'\i sica, Departamento de Fisica Teorica, Modulo C-XI, Universidad Aut\'onoma de Madrid, Cantoblanco E-280049, Spain\\
  $^3$Racah Institute of Physics, The Hebrew University of Jerusalem, Givat Ram, 91904, Israel
  }
\date{Accepted 2010 September 28. Received 2010 September 26; in original form 2010 March 01
}


\maketitle \begin{abstract} \vspace{1pt}
Using a high resolution dark matter (DM) simulation of the Local Group, conducted within the framework of the Constrained Local UniversE Simulation (CLUES) project, we investigate the nature of how satellites of the Milky Way (MW) and M31 are accreted. Satellites of these two galaxies are accreted anisotropically onto the main halos, entering the virial radius of their hosts, from specific ``spots'' with respect to the large scale structure. Furthermore, the material which is tidally stripped from these accreted satellites is also, at  $z=0$, distributed anisotropically and is characterized by an ellipsoidal sub-volume embedded in the halo. The angular pattern created by the locus of satellite infall points and the projected $z=0$ stripped dark matter stripped is investigated within a coordinate system determined by the location of the Local Group companion and the simulated Virgo cluster across concentric shells ranging from $0.1$ to $5 r_{\rm vir}$. Remarkably, the principal axis of the ellipsoidal sub-volume shows a coherent alignment extending from well within the halo to a few $r_{\rm vir}$. A spherical harmonics transform applied to the angular distributions confirms the visual impression: namely, the angular distributions of both the satellites entry points and stripped DM for both halos is dominated by the $l=2$ quadrupole term, whose major principal axis is approximately aligned across the shells considered. It follows that the outer ($r>0.5r_{\rm vir}$) structure of the main halos of the Local Group composed of stripped material is closely related to the cosmic web, within which it is embedded. Given the very plausible hypothesis that  an important fraction of the stellar halo of the Milky Way has been accreted from satellite galaxies, the present results can be directly applied to the stellar halo of the MW and M31. We predict that the remnants of tidally stripped satellites should be embedded in streams of material composed of dark matter and stars. The present  results can therefore shed light on the existence of satellites embedded within larger streams of matter, such as the Segue 2 satellite.

\end{abstract}

\section{Introduction}
\label{introduction} 
In the $\Lambda$CDM model, Galaxy halos like that encompassing our own Milky Way, are formed through the accretion of smaller substructures \citep[e.g.][]{1978MNRAS.183..341W}. Numerical simulations of the formation of galaxy sized halos have consistently shown that small, self bound, over-dense clumps of matter can survive the violent physical processes associated with tidal stripping and dynamical friction and end up as distinct structures embedded within larger halos \citep{1998MNRAS.300..146G,1999ApJ...522...82K,1999ApJ...524L..19M,2000ApJ...539..517B, 2001MNRAS.328..726S,2002MNRAS.335L..84S,2004MNRAS.355..819G,2004MNRAS.351..410G,2010arXiv1002.4200R}. Once accreted, some of these so-called substructures are naturally identified as the sites of satellite galaxies and their properties can be thus be compared to observations of satellites in our own halo.

The satellites of the Milky Way's halo posit a number of interesting questions. That they are all aligned on a great circle in the sky has long been curious \citep{1976RGOB..182..241K,1982Obs...102..202L}. Yet recently \cite{2005A&A...431..517K} argued that the anisotropic distribution of the Milky Way's 11 ``classical'' satellites is inconsistent with the hypothesis that luminous satellites are a random sub-sample of dark matter substructures. Furthermore the newly discovered Sloan Digital Sky Survey (SDSS)  satellites \citep[e.g. see][and references therein]{2008ApJ...686..279K} appear to be located directly on this great disc-of-satellites \citep{2009MNRAS.394.2223M}. 

Many studies \citep{2005MNRAS.363..146L,2007MNRAS.374...16L,2009MNRAS.399..550L,2005ApJ...629..219Z,2005A&A...437..383K} have employed numerical simulations in a bid to understand this peculiar geometry. \cite{2005MNRAS.363..146L} found that the flattened anisotropic distribution of satellites is readily reproduced if the luminous satellites of the Milky Way are associated with substructures that had the largest (pre-infall) progenitors. These subhaloes were born in the dense spines of the filament that collapsed to form the aspherical galactic halo. \cite{2005ApJ...629..219Z} showed that the subhalo population (albeit in Galactic haloes simulated to sit within filaments) as a whole is not isotropically distributed but rather tend to occupy sites close to the long axis of the parent halo. They thus argued that an isotropic subhalo population is not the correct null-hypothesis for the $\Lambda$CDM model

While the planarity of Milky Way satellites is no longer deemed a threat to the standard model, its origin has evaded a definitive understanding. Most authors have indeed managed to reproduce the observed anisotropy and have only hinted at what could be its cause. One of the most interesting ideas is that the satellites of the Milky Way entered the halo as a group. \cite{2008MNRAS.385.1365L} analyzed a suite of high resolution dark matter only simulations populated with galaxies semi-analytically and found that around one third of all satellites were accreted as part of larger groups of satellites. This result is promising since it implies a flattened $z=0$ distribution of satellites. However in constrained simulation of the Local Group, \cite{2009arXiv0909.1916K} found that although satellites form in groups, they tend to be loosely bound, and the group identity is short lived.

A complimentary explanation of the flattened distribution of satellites is that subhaloes are not uniformly accreted from all directions in the sky. Instead, substructures stream across intergalactic space through filaments that feed the growth of a Galaxy-sized halo which has been explicitly shown for Cluster-sized haloes by \cite{2004ApJ...603....7K}. Yet often the effect of the large scale environment is ignored in studies of the spatial distribution of satellite galaxies. 

In this study we employ simulations with constrained initial conditions\footnote{See http://www.CLUES-project.org for a summary of the Constrained Local UniversE Simulation Project.} that aim at reproducing the particulars of the observed Local Group in order to study if substructures enter the halo with a uniform distribution with respect to the large scale structure or whether they come from preferred directions. We examine the effect preferred infall has on the stripped material that is brought in by accreted subhalos.

This paper is organized as follows: in Sec.~\ref{sec:methods} we introduce our simulation, the subhalo finder that was used, and the large scale geometry. In Sec.~\ref{sec:results} we present our results and we conclude in Sec.~\ref{sec:conclusion}

\section{Methods}
\label{sec:methods}
In this section we describe in brief the simulations used as well as the halo and subhalo finding algorithm employed to identify satellites.

\subsection{Constrained Simulations of the Local Group}
The simulations used in this work are embedded in the Constrained Local UniversE Simulation (CLUES) project and have been already studied in a number of recent papers \citep[e.g][]{2009MNRAS.tmp.1707L,2009arXiv0909.1916K,Knebe09Inprep}. We refer the reader to those papers \citep[in particular][]{2009MNRAS.tmp.1707L} for details on how the constraints were generated and how the simulations were run: we highlight just the salient points here for clarity.

We choose to run our simulations using standard $\Lambda$CDM initial conditions, that assume a Wilkinson Microwave Anisotropy Probe 3 cosmology \citep{2007ApJS..170..377S} , i.e. $\Omega_{\rm m} =0.24$, $\Omega_{\rm b} = 0.042$, $\Omega_{\Lambda} = 0.76$ and $h=0.73$. We use a normalization of 
$\sigma_{8} = 0.73$ and an $n=0.95$ slope of the power spectrum. We used the MPI code \textsc{gadget2} \citep{2005MNRAS.364.1105S} to simulate the evolution oft a cosmological box with side length of $L_{\rm box} = 64 h^{-1} Mpc$.

Instead of seeding our initial conditions as a just a random cube of space, the initial conditions of our volume are constrained to reproduce, at $z=0$ a number of objects that compose the local environment \citep[see][for details on how constraints initial conditions are generated]{1991ApJ...380L...5H}, including a ``virgo'' cluster, a ``coma'' cluster and a ``Local Group''. Our method allows us to properly constrain the large scales (i.e. those still linear by $z=0$) but we do not constrain the local group itself. In order to obtain a local group in the correct environment, three low resolutions constrained simulations are run with varying random seeds. Each $z=0$ low resolution simulations is then examined and if an object that resembles the local group is found (by construction this will be in the correct place), these initial conditions are selected for high resolution re-simulation. 

We resimulate just the region of interest around the local group. We center a sphere of radius $2~h^{-1} \rm Mpc$ around the local group and populate with $\sim 5.2\times10^{7}$ low mass, high resolution particles. Within our local group we are thus able to achieve a particle mass of just $M_{\rm dm} = 2.54  \times 10^{5}~h^{-1}\rm M_{\odot}$.

Our constraints reproduce a cosmography which closely resembles the observed Local Group. In Table.~\ref{table:cosm} we compare properties of the simulated local group with observations of the real one\footnote{In a future paper we intend to study in detail the cosmography produced by our constrained simulations.}. Although our results do not match the observations perfectly, the cosmography simulated using our constraints captures the essence - in terms of mass and distances - of the observed Local Group.

\begin{table}
\begin{center}
 \begin{tabular}{l l l l}
Property &Simulated LG  & Observed LG & Reference\\
   \hline
   \hline
$M_{\rm MW}$  & $6.57\times10^{11} M_{\odot}$ & $ 10^{12} M_{\odot} $& [1,2,3] \\
$M_{\rm M31}$  & $8.17\times10^{11} M_{\odot}$ & $8.2\times10^{11} M_{\odot}$ & [4]\\
$M_{\rm Virgo}$  & $1.08\times10^{15} M_{\odot}$ & $1.2\times10^{15} M_{\odot}$ & [5] \\
$d_{\rm MW - M31}$ & 0.909 Mpc & 0.77 Mpc & [6,7] \\
$d_{\rm MW - Virgo}$ & 15.1 Mpc & 16.5 Mpc & [8]\\
$V_{\rm M31}$ & 142 kms$^{-1}$& 195 kms$^{-1}$ & [9]\\
    \hline

 \end{tabular}
 \end{center}
\caption{The $z=0$ properties of the simulated and observed Local Group. From the top row own, we shoe the following properties: the mass of the MW's halo ($M_{\rm MW}$), the mass of M31's halo ($M_{\rm M31}$), the mass of the Virgo cluster ($M_{\rm Virgo}$), the distance from the MW to M31 ($d_{\rm MW-M31}$), the distance from the  MW to Virgo ($d_{\rm MW-Virgo}$), and the relative velocity of M31($V_{\rm M31}$). The references are as follows: [1]~\citet{2008ApJ...684.1143X}; [2]~\citet{2002ApJ...573..597K}; [3]~\citet{2007MNRAS.379..755S}; [4]~\citet{2008MNRAS.389.1911S};[5]~\citet{2001A&A...375..770F}; [6]~\citet{2004AJ....127.2031K}; [7]~\citet{2010A&A...509A..70V}; [8]~\citet{2007ApJ...655..144M}[9]~\citet{2008ApJ...678..187V}}
\label{table:cosm}
\end{table}

\subsection{The halo and subhalo finding algorithm}
In this section, we explain how our halo and subhalo finding algorithm works.
In order to identify halos and subhaloes in our simulation we have run the
MPI+OpenMP hybrid halo finder \texttt{AHF} (\texttt{AMIGA} halo finder, to be
downloaded freely from \texttt{http://popia.ft.uam.es/AMIGA}) described in
detail in \cite{2009ApJS..182..608K}. \texttt{AHF} is an improvement of the
\texttt{MHF} halo finder \citep{2004MNRAS.351..399G}, which locates local
over-densities in an adaptively smoothed density field as prospective halo
centers. The local potential minima are computed for each of these density
peaks and the gravitationally bound particles are determined. Only peaks with
at least 20 bound particles are considered as haloes and retained for further
analysis. In practice for this work, we only consider subhalos with more than 100 particles. We would like to stress that our halo finding algorithm automatically
identifies haloes, sub-haloes, sub-subhaloes, etc. For more details on the
mode of operation and actual functionality we refer the reader to the
code description paper \citep{2009ApJS..182..608K}. 

For each halo, we compute the virial radius $r_{\rm vir}$, that is the radius
$r$ at which the density $M(<r)/(4\pi r^3/3)$ drops below $\Delta_{\rm
  vir}\rho_{\rm back}$. Here $\rho_{\rm back}$ is the cosmological background
matter density. The threshold $\Delta_{\rm vir}$ is computed using the spherical
top-hat collapse model and is a function of both cosmological model and
time. For the cosmology that we are using, $\Delta_{\rm vir}=355$ at $z=0$.  

Subhaloes are defined as haloes which lie within the virial radius of a more
massive halo, the so-called host halo. As subhaloes are embedded within the
density of their respective host halo, their own density profile usually
shows a characteristic upturn at a radius $r_t \lsim r_{\rm vir}$, where
$r_{\rm vir}$ would be their actual (virial) radius if they were found in
isolation.\footnote{Please note that the actual density profile of subhaloes
  after the removal of the host's background drops faster than for isolated
  haloes \citep[e.g.][]{2004ApJ...608..663K}; only when measured within the
  background still present will we find the characteristic upturn used here to
  define the truncation radius $r_t$.}  We use this ``truncation radius''
$r_t$ as the outer edge of the subhalo and hence subhalo properties
(i.e. mass, density profile, velocity dispersion, rotation curve) are
calculated using the gravitationally bound particles inside the truncation
radius $r_t$. For a host halo we calculate properties using the virial radius $r_{\rm vir}$.

We build merger trees by cross-correlating haloes in consecutive simulation
outputs. For this purpose, we use a tool that comes with the \texttt{AHF}
package and is called \texttt{MergerTree}. As the name suggests, it serves the
purpose of identifying corresponding objects in the same simulation at different
redshifts. We follow each halo (either host or subhalo) identified at redshift
$z=0$ backwards in time, identifying as the main progenitor (at the previous
redshift) the halo that both shares the most particles with the present halo \textit{and} is
closest in mass. The latter criterion is important for subhaloes given that all their
particles are also typically bound to the host halo, which is typically orders
of magnitude more massive. Given the capabilities of our halo
finder \texttt{AHF} and the appropriate construction of a merger tree,
subhaloes will be followed correctly along their orbits within the
environment of their respective host until the point where they either are
tidally destroyed or directly merge with the host.

In practice, we use a lower particle limit of 100 particles when considering subhalos. In order to ensure that our results are not resolution dependent, we have also run our analysis using only subhalos with $> 500$ particles. The results are fully consistent with those presented here, albeit with the disadvantage of considering far fewer subhalos. Our particle limit constrains us to only study the haloes history after $z\approx 2$, which corresponds to more than 10 Gyrs ago. This is because before this redshift, the majority of accreted objects fall below our particle limit. Lowering the particle limit used to consider subhalos would have the added advantage of allowing us to probe earlier times, but would return subhalos that were not fully resolved. Thus in the rest of the text when we refer to the``entire history of the halo'', we mean since $z\approx 2$. Furthermore, by $z=2$ the MW and M31 halos have assembled just 7\% and 5\% of their final mass respectively, indicating that the $z<2$ time period corresponds to the main formation epoch of our halos and that the period $z>2$ is negligible from a mass accretion perspective.

\section{Results}
\label{sec:results}
\begin{figure*}
\includegraphics[width=20pc]{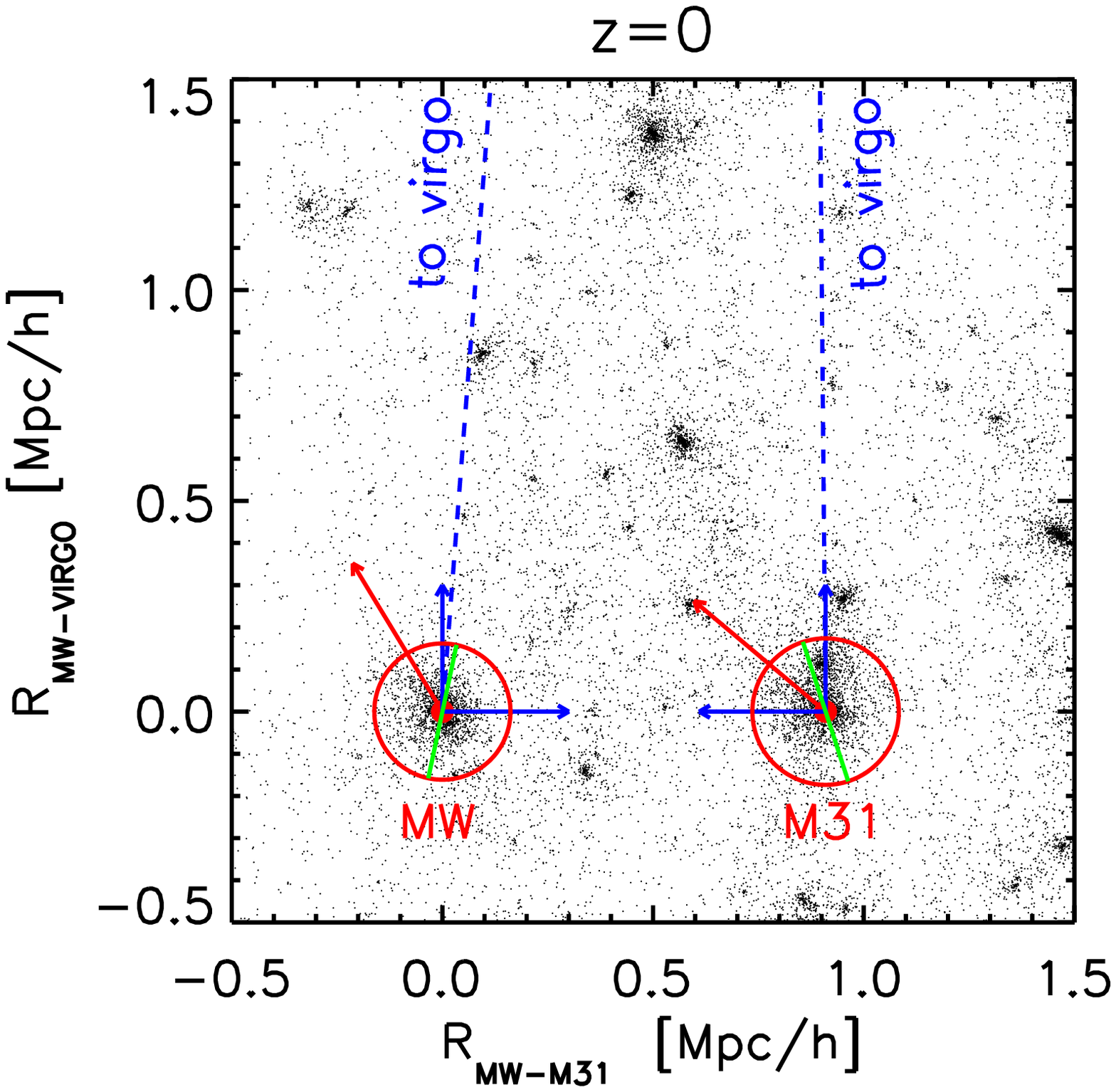}
\includegraphics[width=20pc]{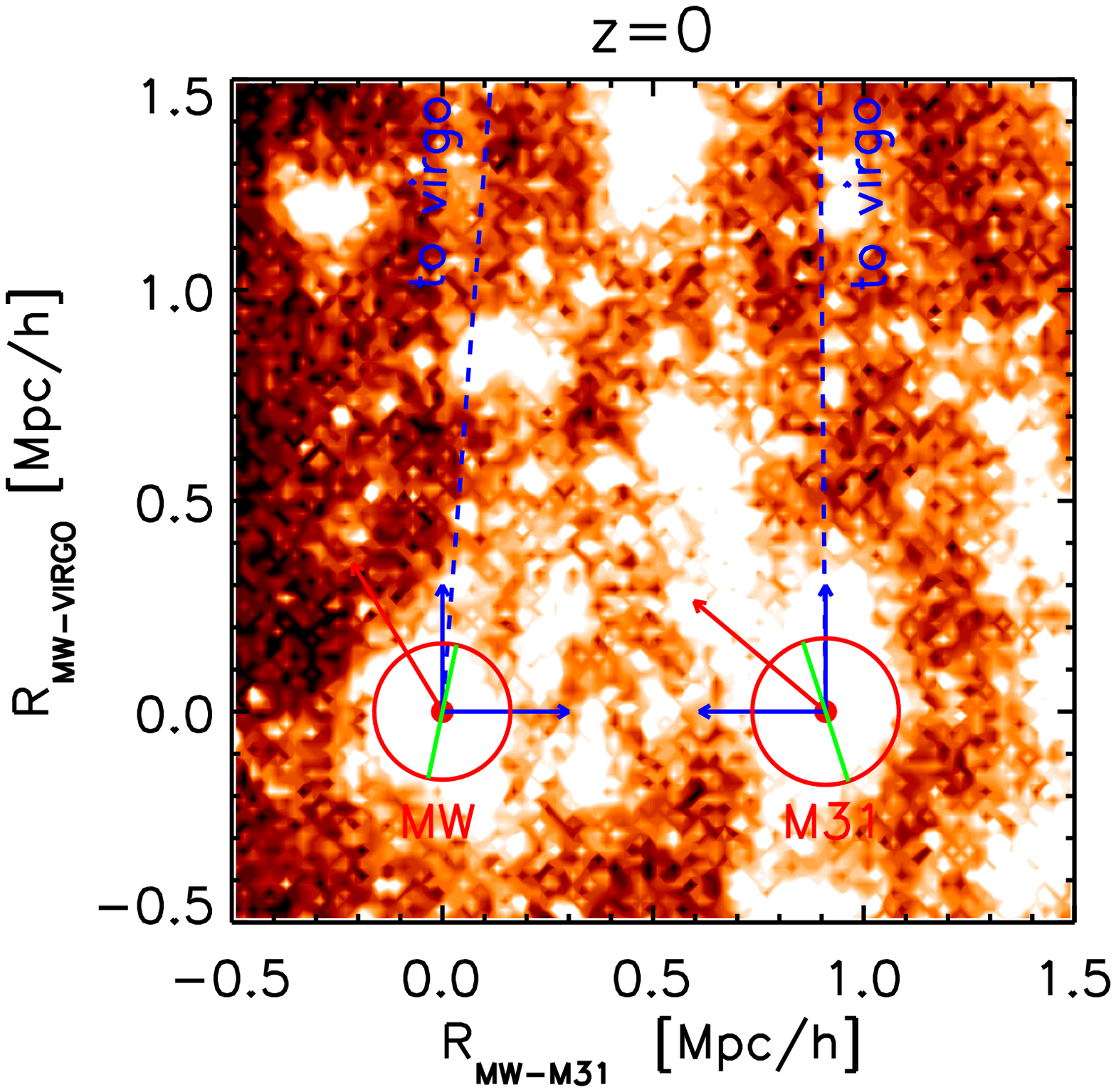}
\caption{A 2$h^{-1}$~Mpc, $z=0$ slice through our simulation showing the coordinate system employed. We have rotated the simulation into the MW's reference frame. The blue arrows show the coordinate axes used, while the dashed blue line shows the direction to the Virgo cluster, some 10 Mpc away. The red arrows (not to scale) indicate the relative velocity of each halo, while the red circles indicate the virial radius (to scale). The green lines indicate the preferential infall direction of satellites across $r_{\rm vir}$ as determined by the eigenvector corresponding to the largest eigenvalue of the quadrupole tensor (see sec.~\ref{sec:shdecomp}). In the panel on the left, the black dots correspond to a random 0.1\% of the dark matter particles. On the right we show a logarithmic density map (linearly contoured) of all the particles in the volume. White regions correspond to high projected densities while red and black regions correspond to lower density regions. Note that the green lines are roughly aligned in the same direction as the nearly-continuous white region, a filamentary like structure.}
\label{fig:coord}
\end{figure*}

In this section we present our results regarding the preferential infall direction of subhalos.  We begin by describing the coordinate system we use in order to fix the axes from which we measure the infalling direction. This coordinate system serves as the basis for determining the preferred direction of accretion. 

\subsection{The Milky Way - Andromeda - Virgo coordinate system}  
Since our constraints accurately simulate the local environment, we define a coordinate system which is fixed with respect to the large scale structure. We show this coordinates system in Fig.~\ref{fig:coord}. From our Merger tree, we know the location of the two halos we choose to call the ``Milky Way'' (MW) and ``Andromeda'' (M31) \citep[e.g. see][]{2009MNRAS.tmp.1707L,2009arXiv0909.1916K} at each snapshot. Additionally, from our constraints, we know the location of the ``virgo'' cluster. We thus define at each snap shot a coordinate system centered on each of these galaxies. We call $R_{\rm MW-M31}$, the position vector connecting MW to M31 (and vice versa) the ``$x$''-axis. Since the two galaxies move with roughly the same velocity in roughly the same direction, this vector changes little over time. We then examine $R_{\rm MW-Virgo}$ and $R_{\rm M31-Virgo}$, the position vector of the Virgo cluster (some 10 Mpc away) centered on the MW and M31. We dot $R_{\rm MW-Virgo}$ and $R_{\rm M31-Virgo}$ with $R_{\rm MW-M31}$ in order to obtain the component of the $R_{\rm MW-Virgo}$ that is perpendicular to $R_{\rm MW-M31}$, calling this vector our ``$y$''-axis. In practice, this vector is (co-incidentally) nearly 90 degrees away from $R_{\rm MW-M31}$ and so requires little adjustment. Once we have an $x$ and $y$ axis, we compute a $z$ axis by crossing $x$ with $y$ (i.e. $z=x \times y$).

The coordinate system, shown in Fig.~\ref{fig:coord}, provides us with three axes that at each time-step are recalculated as the systems evolves. In practice, the coordinate system changes little, as both M31 and the MW move with similar velocities. We may use these axes to note the location on the sky each and every time a subhalo crosses the virial radius or factor thereof.

\subsection{Prefered directions - a qualitative description}
\label{sec:satinfall}
In this section we present a qualitative description of our results regarding the preferred directions we have found. We describe the infall of satellites, the unique $z=0$ distribution of dark matter stripped from these satellites and the self similar distribution across a large radial extent. In practice in order to do identify the point of satellite infall, we examine two adjacent snapshots. We then ask the question: of those subhalos within some radius at the later snapshot, which had progenitors that were outside of this radius one time step earlier? Once all the accreted subhalos have been determined, we identify the point of entry. 
\subsubsection{The preferential infall of satellite galaxies}
In this section we focus on the locations on the sky from which subhalos are accreted. We defer to section~\ref{sec:quant} a more rigorous, quantitative analysis.

\begin{figure*}
\begin{center}
\includegraphics[width=40pc]{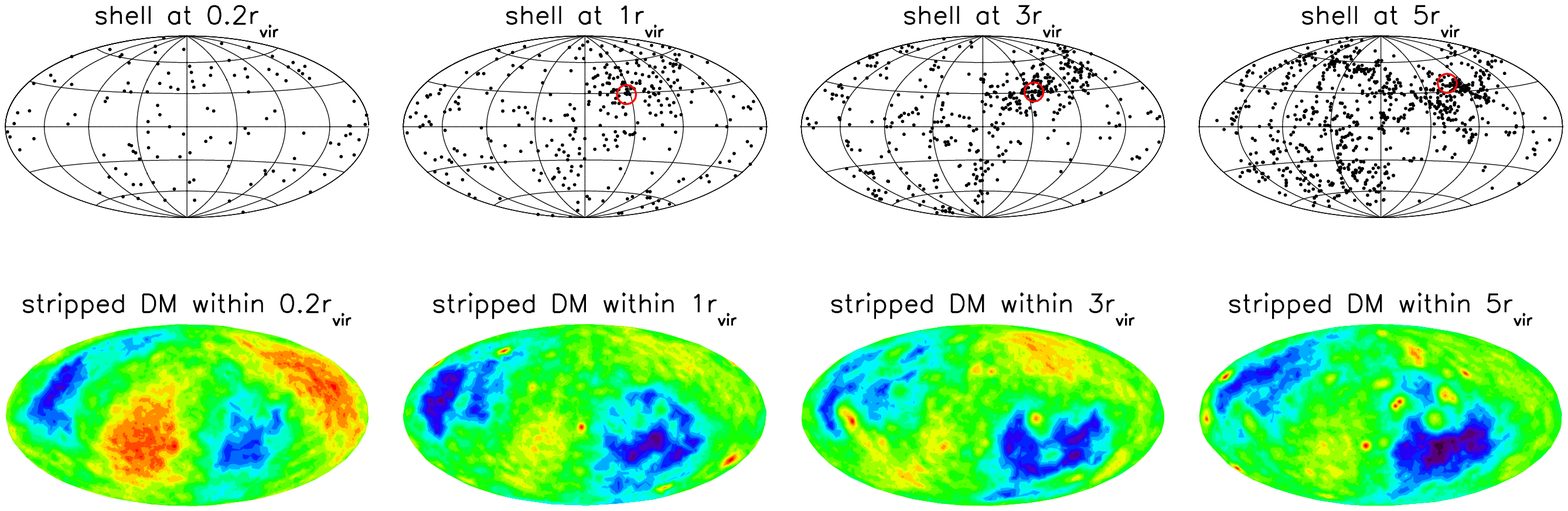}
\includegraphics[width=40pc]{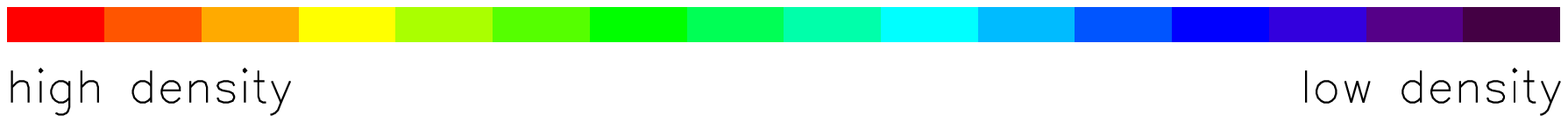}
\end{center}
\caption{\textit{Top Panels:} The point of satellite infall on concentric shells of radius 0.2, 1, 3 and 5$r_{\rm vir}$. Each time a satellite is accreted across one of these shells, its entry point is recorded and plotted.The red circle shows the ``single'' preferred direction, defined as the point on the sky with largest number of satellite infall points with 20$^{\circ}$ of it. \textit{Bottom Panels:} The projected dark matter density of particles brought in by satellites crossing these four radii integrated over the lifetime of the halo. Red regions indicate high density while blue regions indicate low density. Regions on the sky where many satellites entered a given shell,  correspond to regions of higher projected dark matter density.}
\label{fig:satsinfall}
\end{figure*}

In the top panel of Fig.~\ref{fig:satsinfall} we examine the anisotropic accretion of satellites across concentric shells centered on the MW. In the interest of space and focus for all the Aitoff plots in this paper, we omit the same plots for M31 but emphasize that the results are qualitatively and quantitatively very similar. In an Aitoff projection of the sky, we show the location of each satellite's infall point integrated over the entire history of the halo, rotated into our large scale structure coordinate system. In practice, we start by first defining four concentric shells at $r_{c}=$ 0.2, 1, 3 and 5~$r_{\rm vir}$. We then examine two chronologically sequential snapshots and identify all the (sub)halos which were accreted across each spherical surface, plotting them in the upper panels of Fig.~\ref{fig:satsinfall}.

We note that neither the MW (shown here) nor M31 (not shown) display a uniform distribution subhalo accretion points. Satellite infall points are clearly clustered, rather than being uniformly distributed on the sky.

This is our first result: \textit{satellites are not accreted by halos uniformly over the virial radius; rather, they enter the halo from preferred directions.} In order to determine whether this preferred direction is simply dominated by many subhalos entering the halo as a group over a relatively short period of cosmic time \citep[e.g.][]{2008MNRAS.385.1365L}, we have examined the locus of entry points of satellites crossing the virial radius as a function of epoch. Each epoch shows a very similar entry pattern. Simply put: at each epoch in the halo's history, subhalos have a higher tendency to come from one specific direction than from many.

Remarkably, this preferred direction is similar for halos crossing shells from $r_{\rm c} > 0.2~r_{\rm vir}$ to $r_{c}=5~r_{\rm vir}$, reflective of the coherent radial infall across this large range\footnote{Although the $0.5~r_{\rm vir}$ shell is  not shown here, the self similar in-fall pattern begins becoming discernible at $r\approx 0.5~r_{\rm vir}$.}. There are many ways we can identify preferred directions in our distributions. In Sec.~\ref{sec:quant} we perform a more sophisticated analysis on the angular distribution of infall points. Here we simply qualify our visual impression by defining a ``single'' preferred direction, as that infall point with the highest number of other infall points within 20$^{\circ}$ of it. This point is indicated in the upper panels of Fig.~\ref{fig:satsinfall} as the red circle. We also plot it in Fig.~\ref{fig:coord} as the green arrows. We omit this for the left most panel (0.2$r_{\rm vir}$), since the distribution is consistent with random, the preferred entry point is meaningless. For the other three shells at 1, 3 and 5~$r_{\rm vir}$ the distributions are highly clustered in a patch on the sky whose size is roughly $20^{\circ} \times 20^{\circ}$ and at a longitude of between $\theta\sim 50^{\circ}-70^{\circ}$ and a latitude between $\phi \sim 30^{\circ}-50^{\circ}$. We note however, that this naive approach tells us little of the strength of the clustering. 
Although the filamentary nature of the cosmic web can be invoked to explain this affect on large scales, 
its is remarkable that the inflow pattern persists to $r\approx0.5r_{\rm vir}$ (though not shown here).  
Note that the preferred directions changes little across 5~$r_{\rm vir}$.  We explore the self-similar nature of satellite accretion across many scales in section~\ref{sec:selfsimilar}.

It is not directly clear from the simulations what determines this preferred point of entry, although some observations can be made. For one, the subhalos tend to originate from roughly the same region in the sky for both of the two halos: a region in the direction of Virgo. This can be seen in Fig~\ref{fig:coord} as the green arrows. In Fig.~\ref{fig:coord} right panel, we show a logarithmic density map of the volume under consideration, in order to highlight the $z=0$ DM particle distribution. Note that the green arrows can be seen to point in the direction of a finger like, filamentary structure. It becomes apparent from this figure, that as the MW and M31 travel towards virgo they scoop up subhalos in their path. These subhalos reside in a filamentary like structure that also points in the direction of Virgo. As the main halos approach a subhalo, the subhalo comes to a ``fork in the road'' and drifts (perhaps randomly) towards one of the parent halos, later to be accreted. Identifying the filament in the surrounding matter is however, beyond the scope of this paper.

\subsubsection{The $z=0$ distribution of dark matter stripped from accreted satellites}
\label{sec:z0dma_sats}
In this section we look at the $z=0$ distribution of dark matter stripped from the preferentially accreted satellites described above.

We wish to see if there is any hint of the preferential infall of satellites, ingrained in the $z=0$ distribution of dark matter particles. With the exception of the largest subhaloes and those accreted at low $z$, many accreted subhalos will be tidally stripped \cite[e.g.][]{Johnston.98,Penarrubia.etal.06,2008MNRAS.385.1859W} and by $z=0$ will have lost a large amount of their material.  We can identify this material by noting, at each accretion event, which dark matter particles are bound to which subhalo. We then excise from this list, the particles which at $z=0$ are still bound to clumps. We thus obtain a list of dark matter particles that were accreted in subhalos and subsequently stripped from them, over the entire history of the halo. This list can then be used to locate the dark matter particle's $z=0$ positions.

For each halo that is accreted across a given radius ($r_{c}$), we begin by noting the ID's of its dark matter particles. We then locate these particles at $z=0$ and consider only those within $r_{c}$. Using the reference frame defined by the large scale structure at $z=0$ we then project this set of particles onto the surface of a unit sphere and calculate a ``density'' for each particle as the (angular) distance to the 5th nearest neighbor.

We show our results in the bottom panels of Fig.~\ref{fig:satsinfall} for satellites crossing and stripped dark matter particles within shells of radius, $r_{c}=$ 0.2, 1, 3, and 5$r_{\rm vir}$.  A number of salient points can be made from this figure. 

Firstly, by comparing the top and bottom panels of Fig.~\ref{fig:satsinfall}, there appears to be a strong correspondence between the $z=0$ angular distribution of stripped dark matter and the locus of satellite entry points. Although the correspondence is not one to one, regions from where many satellites entered the virial radius coincide with regions of high stripped particle density. Likewise, regions where few satellites entered the halo have correspondingly fewer dark matter particles. This brings us to our second main result: \textit{dark matter that is stripped from subhalos, displays at $z=0$ a memory of the region from where the parent subhalo was accreted}.  We therefore can infer the following picture of subhalo infall: subhalos are accreted from a preferred direction over the history of the halo. As these subhalos move through the halo, they suffer mass loss effects due to tidal forces. Their constituent particles get stripped and the subhalos - at some point - cease to exist as recognizable over dense clumps. The stripped material however, continues to reside in regions centered on where the subhalo was accreted.

We note that stripped debris material is a sub-set of halo particles not bound to substructures. The distribution of the latter is mildly aspherical while the debris material appears significantly flattened.

We wish to draw the reader's attention to the self-similar patterns across the three shells (where $r_{c} > 0.2~r_{\rm vir}$) shown in Fig.~\ref{fig:satsinfall}. A visual inspection immediately reveals that the preferred direction on the sky from where satellites are accreted changes little from 1 to 5~$r_{\rm vir}$. Similarly, the $z=0$ dark matter brought in by these satellites also occupies similar regions of the sky. Broadly speaking, \textit{regions of high satellite infall point density correspond to regions of high $z=0$ stripped dark matter density and these regions are constant across an order of magnitude in radial distance}.

We would like to remind the reader that the projected angular surface density maps in (the bottom panel of) Fig.~2 are cumulative, i.e. all the stripped material within the respective radius is considered. This means that some particles may (or may not) contribute to more then one map. Consider a subhalo that crosses both 5 and 3 $r_{\rm vir}$ yet only deposits its stripped material interior to 3~$r_{\rm vir}$: its stripped material will contribute to the density map within both 3$r_{\rm vir}$ and 5$r_{\rm vir}$ hence possibly enhancing the signal of alignment. To gauge the relevance of this "particle stacking" we examined surface density maps considering just the stripped material that at $z=0$ is located within the concentric shells probing the regions: $0.2r_{\rm vir}\leq r \leq r_{\rm vir}$, $r_{\rm vir}\leq r \leq 3r_{\rm vir}$, and $3r_{\rm vir}\leq r\leq5r_{\rm vir}$. We qualitatively recover the same signal that shows the same characteristics as the maps presented in the bottom panels of Fig.2, respectively.

\subsubsection{The self-similar angular distribution of stripped material}
\label{sec:selfsimilar}

\begin{figure*}
\begin{center}
\includegraphics[width=40pc]{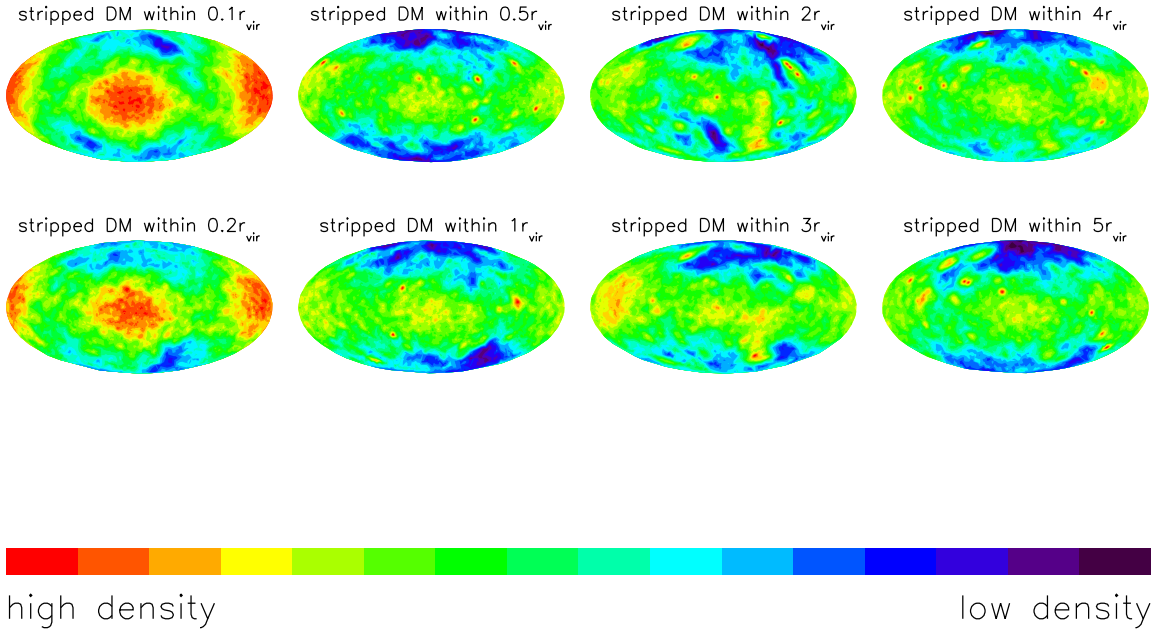}
\includegraphics[width=40pc]{./colourbar.eps}
\end{center}
\caption{The projected dark matter density ellipsoid within 8 concentric shells of radius 0.1, 0.2, 0.5, 1, 2 3, 4 and 5~$r_{\rm vir}$, rotated into the principal axis calculated at 1~$r_{\rm vir}$. Note how the orientation of the  dark matter ellipsoid's over 5~$r_{\rm vir}$ changes little. The dark matter stripped from satellites accreted across 0.1~$r_{\rm vir}$ has an orientation consistent with that material accreted at a distance of 5~$r_{\rm vir}$: a factor of  50 times greater. Note that the (bottom) panels displaying stripped dark matter within 0.2, 1, 3, and 5$r_{\rm vir}$ are simply rotations of the the same material shown in the bottom panels of Fig.2. }
\label{fig:dmz0}
\end{figure*}

In this section we examine in detail the distribution of the dark matter stripped from satellites accreted across concentric shells centered on the galaxy.

In Fig.~\ref{fig:satsinfall} we showed that both the locus of satellite infall points and the dark matter ellipsoid of stripped particles, displays coherence both with each other and across many radial scales. 
We wish to study this ellipsoid in more detail.  In Fig.~\ref{fig:dmz0} we show an Aitoff projection of the $z=0$ angular position of particles stripped from satellites that crossed eight shells at $r_{c} =$ 0.1, 0.2, 0.5, 1, 2, 3, 4, and 5~$r_{\rm vir}$. We rotate each particle distribution into the principal axis defined by these particles within $r _{\rm vir}$. We find the principal axes of the particle distribution by diagonalizing the moment inertia tensor:
\begin{equation}
I_{i,j}=\sum_{n}m_{n}x_{i}x_{j}
\end{equation}
in order to identify the eigenvectors and eigenvalues (defined according to convention as $a>b>c$). Note that these are particles that composed subhalos with more than 100 particles that were accreted since $z\approx2$: they make up roughly $20\%$ of the halo's $z=0$ particles (the other $80\%$ were either added to the halo through ``ambient'' accretion or in subhalos below our particle limit or are still bound to subhalos). Around $75\%$ of these particles have become unbound from their original subhalo due to stripping processes and are, by $z=0$, bound only to the main halo.

We note that unlike the background material, these particles are not uniformly distributed in the halo according to e.g. a \cite{1997ApJ...490..493N} profile - or in fact according to any axis-symmetric profile. Instead they appear to define a thick plane embedded in the roughly spherical material. This is our third result: \textit{particles stripped from subhalos that are accreted from a preferred direction occupy, at $z=0$, a confined ellipsoidal sub-volume embedded within the halo}.  At a given radial distance, the long axis of this ellipsoidal sub-volume is roughly aligned with that of the background dark halo within the same radius.

We wish to emphasize that both the satellite infall pattern (shown in the top panels of Fig.~\ref{fig:satsinfall}) and the $z=0$ stripped dark matter (shown in the bottom panels of Fig.~\ref{fig:satsinfall} and in Fig.~\ref{fig:dmz0}) is remarkable constant, with respect to our coordinate system, across a large radial distance. From Fig.~\ref{fig:dmz0} we note that its is remarkable how little the orientation of stripped dark matter changes over such a huge distance. As mentioned in section~\ref{sec:z0dma_sats} we wish to emphasize that the satellite infall pattern at 5~$r_{\rm vir}$ (roughly 1 Mpc) is coherent with that at distance of $\gsim0.1~r_{\rm vir}$ (roughly 40 kpc). This brings us to our fourth main result: \textit{the angular distribution of satellite infall points on concentric shells centered on the main galaxy, is the same regardless of at what distance the accretion event is measured}. Additionally, \textit{the dark matter ellipsoid composed of particles stripped from these satellites is oriented in the same direction across an order of magnitude in extent}. In other words, the $z=0$ position of material stripped from satellites that crossed 5~$r_{\rm vir}$ is in roughly the same configuration as the $z=0$ material stripped from satellites that penetrated the inner most part of the halo, $0.1~r_{\rm vir}$. Likewise, we can see a correspondence between the angular distribution of satellites that fall into shells of distance $\gsim 0.2~r_{\rm vir}$ as those that enter $5~r_{\rm vir}$. Note however, that there is less material in the inner most shells for two reasons: first it is a much smaller volume, but more importantly, far fewer satellites cross the inner shells. Crossing the inner shells of the halo is rarer since fewer subhalos are on orbits which bring them to the inner parts - \cite{2004MNRAS.351..410G} found that the  pericentre distribution of subhaloes peaks at about $\sim0.35~r_{\rm vir}$. Furthermore, even if a subhalo does penetrate these depths, it has most likely lost enough particles (due to tidal stripping) that it fails to make our mass cut.

\begin{figure}
\begin{center}
\includegraphics[width=20pc]{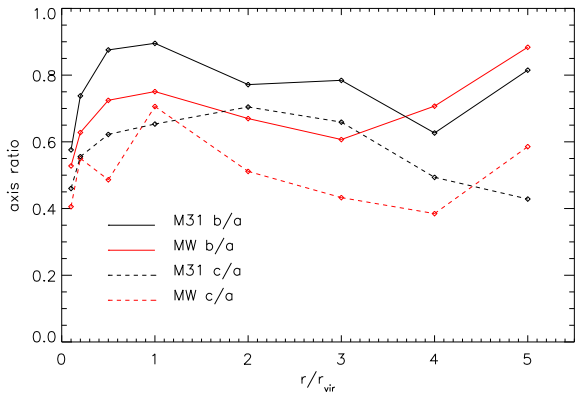}
\end{center}
\caption{The sphericity (as measured by the axes ratios) of the stripped dark matter particle ellipsoid as a function of radius. We show this quantity for the MW halo (red) and for the M31 halo (black). We show the the ratio of the intermediate to long axis ($b/a$) as solid lines and the ratio of the short to long ($c/a$) in as dotted lines.}
\label{fig:flat}
\end{figure}

\begin{figure*}
\begin{center}
\includegraphics[width=40pc]{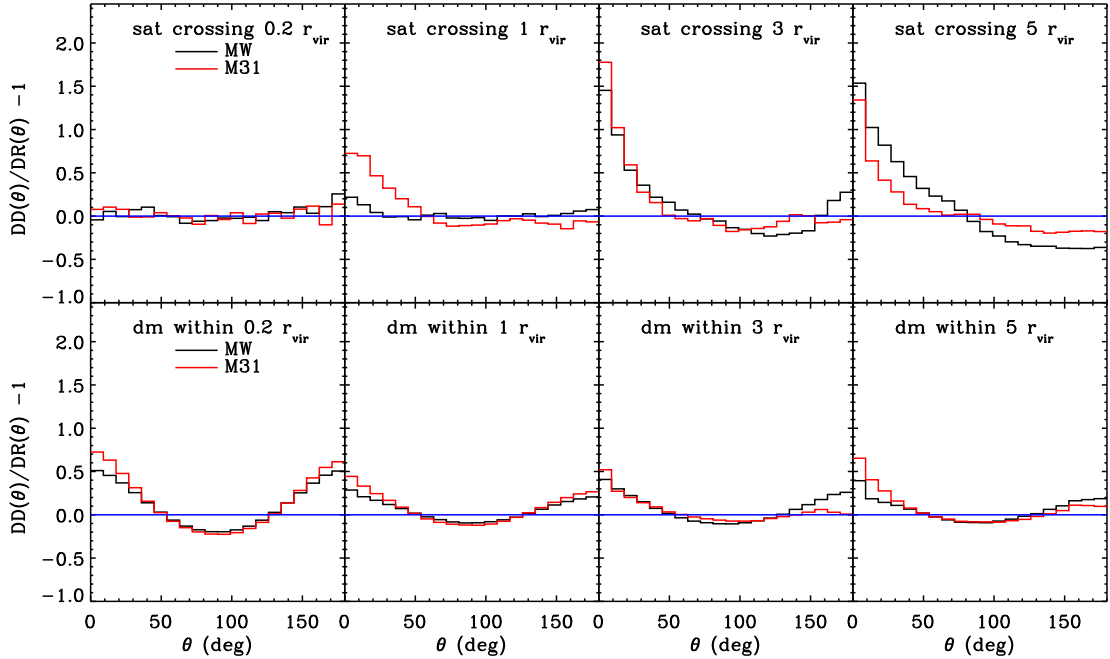}
\end{center}
\caption{The angular auto-angular correlation function for satellite infall points (above) and dark matter stripped from accreted satellites (below) for 4 shells of radius 0.2, 1, 3, and 5~$r_{\rm vir}$ (left to right) for the MW (black) and M31 (red). The blue line indicates a random distribution on the sky.}
\label{fig:acf}
\end{figure*}

The degree of flattening of the two triaxial sub-volumes can be seen in Fig.~\ref{fig:flat} where we plot the axis ratio of the ellipsoid as a function of radius. Note that the axes ratios for both halos at $r=r_{\rm vir}$ indicates that these particles have a triaxial (i.e. $a>b>c$) distribution. Both halos show that the particles stripped from preferentially accreted subhalos inhabit prolate forms within $r_{\rm vir}$, becoming more oblate out to $r_{\rm vir}$. Note however, that in Fig.~\ref{fig:dmz0}, the reference frame is fixed at $r_{\rm vir}$, indicating that even thought the ellipsoid is stretched along the main axis from oblate to prolate, the \textit{principal direction} does not change, regardless of where it is found.

\subsection{Quantifying the preferential infall of satellites and their stripped dark matter}
\label{sec:quant}
In this section we wish to quantify what is visually apparent in Fig.~\ref{fig:satsinfall}; the non uniformity of satellite infall points, the anisotropic distribution of $z=0$ dark matter stripped from preferentially accreted satellites and the self-similar angular distribution of these two components across an order of magnitude in radial extent.
 
\subsubsection{The clustering of satellite infall points and $z=0$ dark matter particles}

We begin to quantify this non-uniformity by computing the (auto) angular-correlation function of the two distributions shown in Fig.~\ref{fig:satsinfall}. For each shell at a given radius, we know the distribution of data points: the satellite infall positions and $z=0$ dark matter particles. We then construct a random sample with the same number of points as the data. For each data point, we calculate the distribution of angles ($\theta$) to all other data points $DD(\theta)$, and to all random points $DR(\theta$). We then construct the auto-correlation function:
\begin{equation}
w(\theta)=\frac{DD(\theta)}{DR(\theta)}-1
\label{eq:acf}
\end{equation}

In Fig.~\ref{fig:acf} we plot the angular correlation function of satellite infall points (top panels) for the same four concentric shells shown in Fig.~\ref{fig:satsinfall} namely,  $r_{c}=$ 0.2, 1, 3 and 5~$r_{\rm vir}$ for the MW (black) and M31 halos (red). Note that a random angular distribution would have $DD(\theta)=DR(\theta)$ and thus be represented in Fig.~\ref{fig:acf} as a horizontal line at null. The top panels of this figure quantify what is visually apparent from Fig.~\ref{fig:satsinfall}: satellites that enter shells greater than the virial radius do so anisotropically - from preferred directions. The angular correlation function of satellite infall points across shells deep within the virial radius, is consistent with random. 

Like the satellite infall points, the stripped dark matter is far from being uniformly distributed in the halo (or on the sky). We show the projected angular correlation function of dark matter particles stripped from accreted subhalos in the bottom panels of Fig~\ref{fig:dmz0} across four concentric shells of radius $r_{c}=$ 0.2, 1, 3, and 5$r_{\rm vir}$ for the MW (black) and M3 halos (red). Within $r_{\rm vir}$ we see a clear bipolar signal indicating a smooth ellipsoidal distribution. As we go to larger radii, the bipolar signal is weakened as the dark matter becomes more clustered. 

We have also performed the following KS test on the distribution of satellite infall points. For each spherical surface we place many random strips on the sky with an angular thickness of 15$^{\circ}$. For each strip we calculate the Kolmogorov-Smirnoff (KS) probability that the distribution contained within it is consistent with a random distribution. We then take the average KS probability of all strips to get a feeling for how random the sample is. We find the same conclusion: within $r_{\rm vir}$ the KS test accepts the hypothesis that satellite infall points are consistent with a random distribution, while outside $r_{\rm vir}$, this hypothesis is rejected at high confidence levels. Performing the same KS test on the stripped dark matter particles indicates an inconstancy with a random sample at high significance levels.

\subsubsection{Spherical Harmonic Decomposition}
\label{sec:shdecomp}
\begin{figure}
\includegraphics[width=20pc]{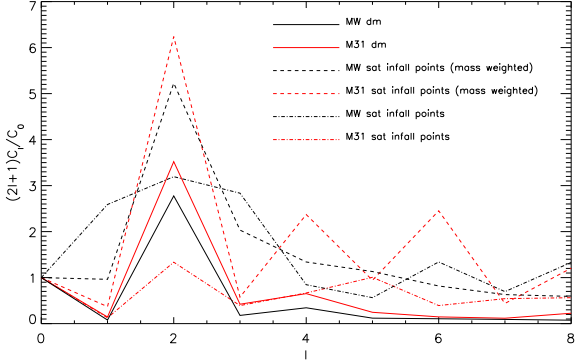}
\caption{The angular power spectrum of the satellite infall points (dot-dashed lines) with a mass weighting (dashed) and stripped DM (solid lines) for the MW (black) and M31 (red) halos at $r_{\rm c}=r_{\rm vir}$. The plot shows the power per harmonic number, $\ell$, normalized by the monopole term, $C_0$.}
\label{fig:powspec}
\end{figure}
\begin{figure*}
\includegraphics[width=40pc]{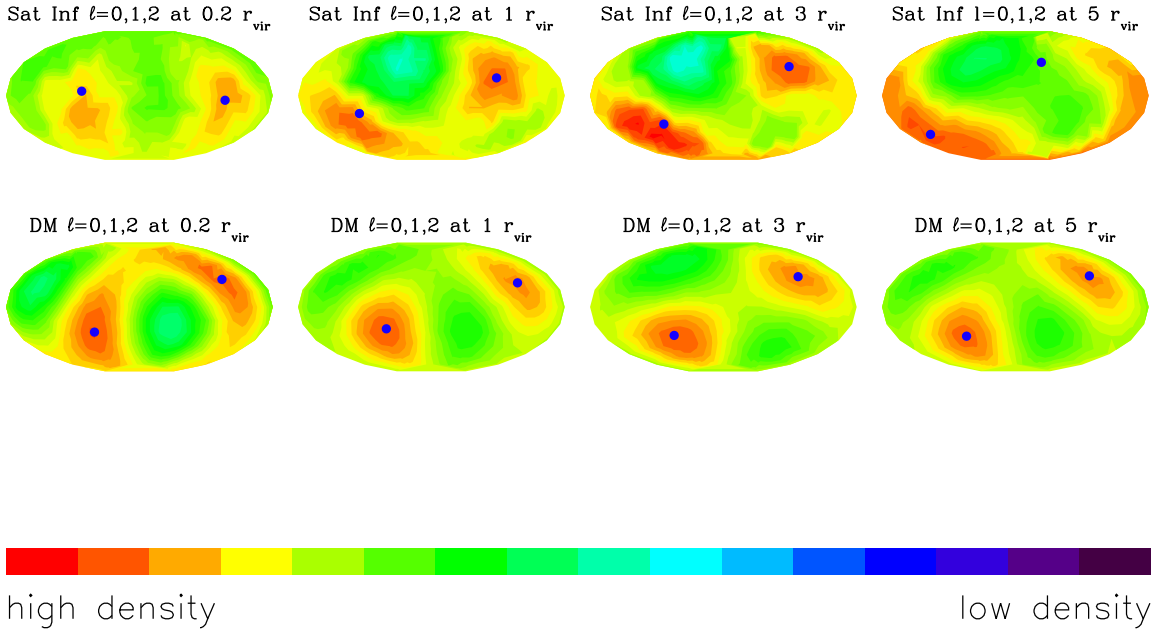}
\includegraphics[width=40pc]{./colourbar.eps}
\caption{The first three harmonics (up to the $\ell=2$ quadrupole) of the spherical harmonic decomposition of the angular distribution of satellite infall points (top) and the $z=0$ dark matter stripped from these satellites (bottom) across our four fiducial shells. The blue dots indicate the direction of the largest eigenvector of the quadrupole tensor.}
\label{fig:shmw}
\end{figure*}

The angular distribution is further investigated here by means of a spherical harmonic decomposition  \cite[e.g.][]{1980lssu.book.....P} of the angular distribution of satellite infall points and stripped dark matter particles. Although technically speaking only the satellite infall points sit on the surface of a sphere, in order to properly deal with the dark matter particle distributions, we  project their positions onto the surface across which their host satellites were accreted.

Consider the angular distribution of $N$ discrete objects. The spherical harmonic transform of the angular distribution of these points is then given by:
\begin{equation}
a_{\ell}^{m}=\frac{1}{N}\sum^{N}_{i=1}Y_{\ell}^{m}(\theta_{i},\phi_{i})
\end{equation}

The power spectrum as a function of spherical harmonic, $\ell$ can be calculated as
\begin{equation}
C_{\ell}=\frac{1}{2\ell+1}\sum^{\ell}_{m=-\ell}|a_{\ell}^{m}|^{2}
\label{eq:powspec}
\end{equation}
The angular power spectrum is important because it tells us which $\ell$ mode contains the most power - or which $\ell$ mode is contributing the most to the spherical harmonic decomposition. 

Given a dark matter particle or satellite infall point with coordinates $(\theta,\phi)$ we may evaluate the spherical harmonic function for a given $\ell$ and $m$. The original discrete point distribution can be approximated by a smooth angular surface density constructed  by summing over all $m$'s and $\ell$'s up to some $\ell_{\rm max}$ according to the following equation:
\begin{equation}
\sigma(\theta,\phi)=\sum_{\ell=0}^{\ell_{\rm max}}\sum_{m=-\ell}^{\ell}a_{\ell}^{m}Y_{\ell}^{m}(\theta,\phi)
\end{equation}

Some of the modes of the spherical harmonic decomposition are particularly interesting. While the $\ell = 0$ monopole term is trivial (in our formulation it is a constant $C_{0}=1/4\pi$), the $\ell =1$ dipole mode and the $\ell = 2$ quadrupole provide us with preferred directions. The relative importance of each mode can be obtained by computing the power spectrum, $C_{\ell}$, according to Eq.~\ref{eq:powspec}.

In Fig.~\ref{fig:powspec} we show the angular power spectra for the $r_{\rm vir}$ shell of the
MW and M31 (black and red lines respectively) calculated using equation~\ref{eq:powspec}. Note that we normalize our angular power spectra by the trivial monopole ($\ell=0$) contribution such that the power at a given $\ell$ is measured relative to the monopole term. The satellite infall angular power spectra are shown as dot-dashed lines while the angular power spectra for dark matter particles stripped from preferentially accreted satellites are shown as solid lines. We highlight that in all four angular power spectra the maximum occurs at $\ell=2$, indicating that the quadrupole moment dominates the angular power spectrum. 

In the case of satellite infall points, we note that although the $\ell =2$ quadrupole moment is the maximum of the power spectrum, other moments are not necessarily negligible. In order to further investigate the importance of other moments, we weigh each spherical harmonic transform by the mass of the accreted satellite, $m_{i}$ and divide by the total mass accreted in all satellites, $M$
\begin{equation}
a_{\ell}^{m}=\frac{1}{N M}\sum^{N}_{i=1}m_{i}Y_{\ell}^{m}(\theta_{i},\phi_{i})
\end{equation}
The mass weighted power spectrum is shown in Fig.~\ref{fig:powspec} as the dashed lines. Weighing each infall point by the mass of the satellite drastically increases the importance of the $\ell=2$ quadrupole term. This implies that although all satellites are being anisotropically accreted, the signal is strongest for the most massive satellites. This makes intuitive sense: the most massive satellites are born in the densest parts of the filament while smaller halos are more widely scattered across the thick filaments, reflecting their weaker clustering strength \citep{1989MNRAS.237.1127C,1999MNRAS.304..175M}.

The dipole (and higher order moments) is relatively weak compared with the quadrupole. We thus focus our analysis on the quadrupole and its principal direction. In particular we can define the (dimensionless and traceless) quadrupole tensor, normalized by the number of particles, by writing
\begin{equation}
\label{eq:quad}
Q_{\alpha,\beta}=\frac{1}{N}\sum_{i=1}^{N}\big(\frac{x_{i}^{\alpha}x_{i}^{\beta}}{r^{2}}-\frac{\delta_{\alpha\beta}}{3}\big)
\end{equation}
where $x^{\alpha}_{i}$ and $x^{\beta}_{i}$ are just the ($x,y,z$) positions of the $i^{th}$ particle. In practice the individual elements of the symmetric quadrupole tensor can be computed directly from the $Y_{\ell}^{m}$'s \citep[e.g. see][]{1962clel.book.....J}. Once we have obtained the quadrupole tensor, we can diagonalize it to obtain the principal eigenvectors which quantify the preferred directions that are visually apparent in. Fig.~\ref{fig:satsinfall}. The eigenvalues are numbered in decreasing ordered such that the first eigenvector corresponds to the largest eigenvalue (hereafter referred to as the ``largest eigenvector'').

In Fig.~\ref{fig:shmw} we show the $\ell=0,1,2$ spherical harmonic decomposition of the satellite infall points (top) and the dark matter particles contained in preferentially accreted satellites (bottom) as a function of infall radius for the MW halo, noting that the M31 halo looks very similar. The two directions (positive and negative) of the largest eigenvector of the diagonalized quadrupole tensor ($\hat{a}_{\rm sat quad}$ for the satellite infall points and $\hat{a}_{\rm dm quad}$ for the dark matter particles) are shown as the two blue dots in each panel. Note that the small misalignment between the two lobes of high spherical angular over density and the directions of the largest eigenvector of the quadrupole tensor, is indicative of the existence of other albeit minor moments.

When looking at the satellite infall panels, we note that the quadrupole dominates out to the final $5~r_{\rm vir}$ panel where it begins to become weaken. Note also that although the direction of the $\hat{a}_{\rm sat~quad}$ changes across $5~r_{\rm vir}$, it does so slowly and incrementally. Furthermore, the location on the sky of $\hat{a}_{\rm sat~quad}$ is close to that of $\hat{a}_{\rm dm~quad}$ which also changes very incrementally across $5~r_{\rm vir}$. Note that the angular distribution reconstructed from the first three harmonics shown in Fig.~\ref{fig:shmw} looks qualitatively similar to Fig.~\ref{fig:satsinfall}. The largest eigenvector of the quadrupole tensor points close to the ``single'' preferred direction (the red circle in Fig.~\ref{fig:satsinfall}) identified as the patch on the sky with the most infall points within it. Had we wanted to fully reconstruct the angular surface density, we would have to consider harmonics beyond $\ell=2$.

\begin{figure}
\includegraphics[width=20pc]{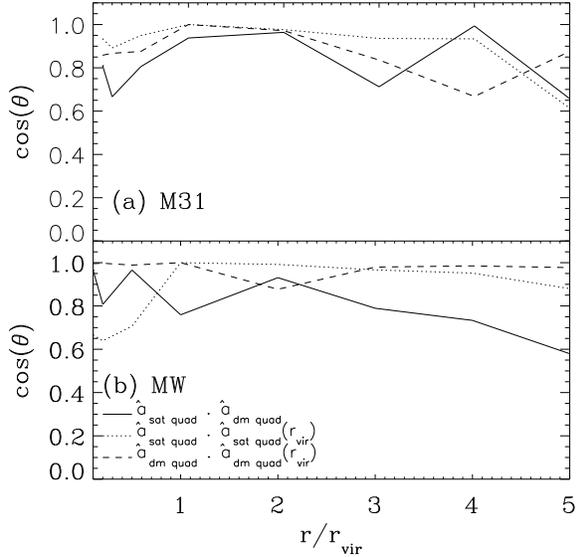}
\caption{The orientation of the first eigenvector ($\hat{a}$) of the $\ell =2$ quadrupole for M31 (top, panel a) and the MW (bottom, panel b). The solid line shows the angle between the largest eigenvector of the satellite infall point distribution $\hat{a}_{\rm sat}$ and that of the dark matter ellipsoid $\hat{a}_{\rm dm}$ as a function of radius. The dotted line shows the value of the angle between $\hat{a}_{\rm sat}$ at a given radius and $\hat{a}_{\rm sat}(r_{\rm vir})$, at the virial radius. The dashed line shows the same quantitiy for the Dark matter ellipsoid, namely the angle between $\hat{a}_{\rm dm }$ and $\hat{a}_{\rm dm}(r_{\rm vir})$.}
\label{fig:quadalign}
\end{figure}

We now quantify the orientation as a function of radius of the largest eigenvectors of the quadrupole $\hat{a}_{\rm sat~quad}$ and $\hat{a}_{\rm dm~quad}$ in Fig.~\ref{fig:quadalign} for the M31 halo (top, panel a) and the MW halo (bottom, panel b). We show the angle between $\hat{a}_{\rm sat~quad}$ and $\hat{a}_{\rm dm~quad}$ as a function of radius as the solid line. For both halos, the cosine of this angle is $\sim 1$ within $r_{\rm vir}$, indicating that the two are well aligned with each other. Beyond $r_{\rm vir}$ the alignment weakens for the MW at  $r> 3~r_{\rm vir}$.  In the MW's case, the weakening of the alignment at greater radii is indicative of a less collimated in-flow. Stripped DM brought at greater radii is spread over a larger angular extent indicating less radial satellite trajectories. For M31, the mild weakening of the alignment  indicates  that the collimation is more robust - satellites fall in on relatively radial trajectories.

We now address how the largest eigenvectors of the $\ell=2$ quadrupole changes with respect to the virial values. The dotted and dashed lines in Fig.~\ref{fig:quadalign} show the (cosine of the) angle between $\hat{a}_{\rm sat~quad}$ and $\hat{a}_{\rm dm~quad}$ and the $\hat{a}_{\rm sat~quad}(r_{\rm vir})$ and $\hat{a}_{\rm dm~quad}(r_{\rm vir})$, the eigenvectors at the virial radius. By definition both these lines lines go through the point (1,1). For the MW halo, (Fig.~\ref{fig:quadalign}a) we see that these are remarkable stable and are well aligned out to $5~r_{\rm vir}$. We see a very similar picture for M31.

The spherical harmonic decomposition presented in the preceding section allows us to quantify the visual impression immediately apparent from Fig.~\ref{fig:satsinfall}. The principal directions are identified as the largest eigenvectors of the $\ell =2$ quadrupole tensor. We see that indeed they point towards the regions of high satellite infall point density and high dark matter projected density. We can thus measure the orientation between the eigenvectors and can determine how they move about with increasing radius.

\section{Summary and Conclusions}
\label{sec:conclusion}
In this paper we have used constrained cosmological simulations of the formation of our Local Group to examine the accretion of satellites galaxies. Our simulations reproduce the essential features of the local group allowing us to draw conclusions that are directly applicable to the observed Local environment.

We have examined the locations on the sky from which satellite galaxies are accreted onto the simulated MW and M31. We use the large scale structure to fix a coordinate system in which we measure the distribution of infall points across a number of radii. We found five important results:
\begin{itemize}
\item Integrated over the entire history of the halo, subhalos do not enter concentric spherical surfaces (including the virial radius) from all directions, rather they tend to be accreted from prefered directions on the sky. 
\item Part of the dark matter bound to these subhalos at accretion gets stripped and becomes bound to the main halo. At $z=0$, however these particles define an ellipsoidal sub-volume embedded within the halo.
\item The orientation of this sub-volume is consistent with the preferred direction of the accreted progenitor subhalos: the preferred direction is perpendicular to the short axis of this ellipsoid. The dark matter stripped from these subhalos thus retains a memory of the prefered direction of infall.
\item The infall pattern across an order of magnitude in radial distance indicates that there is a coherent distribution from roughly 1~Mpc ($5~r_{\rm vir}$) to 20~Kpc ($0.1~r_{\rm vir}$). Satellites cross concentric shells at roughly the same regions on the sky. Also the $z=0$ stripped dark matter in these concentric shells displays a remarkable angular association with the locus of satellites infall points.
\item By performing a spherical harmonic decomposition of the satellite infall points and the $z=0$ dark matter stripped from them, we have identified the $\ell=2$ quadrupole moment as the dominant mode. The satellite infall point quadrupole moment is even more dominant when the spherical harmonic decomposition is weighted by the mass of the satellite.  
\item The first eigenvector of the quadrupole moment, corresponding to the highest eigenvalue, defines a line along which the satellites are preferentially accreted and where at $z=0$ most of the stripped dark matter particles reside. For the satellite infall points, this line points to within $\sim30^{\circ}$ of the patch on the sky with the highest density of satellite infall points. The principal axes calculated for different shells are remarkably well aligned over  a wide radial extent.
\end{itemize}

Our results are important for a number of studies. Firstly, assuming that subhalos harbor luminous galaxies, and that these galaxies can be associated with the tidally stripped dark matter subhalos, our results confirm early works that attempt to explain the origin of a flattened distribution of satellites. Since the stripped material of accreted subhalos defines a flattened distribution at $z=0$, so too should the galaxies they once harbored.

The current analysis is focused on a DM only constrained simulation of the Local Group, focusing on the role of satellites and the DM carried by the satellites into the main halo. This DM component constitutes roughly $20\%$ of the total mass of the halo. Such a relatively small fraction might raise doubts on the significance of our findings on the distribution of that component. However, current understanding of galaxy formation points towards a picture in which the galactic stellar halo is predominantly made of stars formed in the accreted satellites \citep{2008ApJ...680..295B}. Given that stars behaves dynamically in a manner similar to simulated DM particles, it follows that our conclusions concerning the DM carried by satellites pertains equally to the stellar halo. Our results therefore predict that the tidal remnants of stripped satellites (for example, Globular Clusters) should be embedded in of streams dark matter and stars. 

We also find agreement with studies of larger systems (e.g. clusters, \cite[see][]{2004ApJ...603....7K}) which have shown that satellites are accreted by clusters in preferred directions that coincide with filaments. Our results show that accretion of subhalo from preferred directions occurs on the galactic halo mass scale as well. 

Additionally, our results indicate that for galactic halos, there appears to be conherence in the infall patterns. Although only small percentage of satellites are on purely radial orbits, collapsing filaments have well defined directions, giving rise to satellites that can penetrate deep into the halo's inner parts from the same directions from where the filament originated.

An interesting observation can be made regarding to the $z=0$ orientation of the two halos with respect to the larger scales. \cite{2004ApJ...613L..41N} argued that the rotation axis of a central (disc) galaxy aligns itself perpendicular to the minor axis of the surrounding matter. This can perhaps be seen in the real MW, as the North Galactic Pole points almost directly to the Virgo cluster. In our simulations we find a similar set-up: the angular momentum axis of our two halos seem to lie close to the plane defined by our coordinate system and point in the direction of our virgo cluster.

Another interesting observation is the existence of a new satellite population known as  ``satellites of satellites''. Specifically, \cite{2009MNRAS.397.1748B} discovered a new satellite of the Milky Way - Segue 2 - embedded in a stream of matter. \cite{2009MNRAS.397.1748B} concluded that the stream of matter is the remnant of a tidally stripped parent satellite to which Segue-2 once belonged. Since we have found an association between stripped satellite material and the location of satellite infall points, the fact that this object is embedded in a stream of material is readily understandable, if the stream is composed of (dark) matter that has recently been unbound and is moving with the satellite. Our results indicate that many satellites should be embedded in ``streams'' defined as coherent dark matter ellipsoids.

\section*{Acknowledgments}
We thank the anonymous referee for perceptive remarks, which improved the paper. NIL is supported by the Minerva Stiftung of the Max Planck Gesellschaft. AK is
supported by the MICINN through the Ramon y Cajal programme. YH has been partially supported by the ISF (13/08). GY would like to thank the MICINN (Spain)  for financial support under project numbers FPA 2009-08958 and AYA 2009-13875-C03 and the SyeC Consolider project CSD 2007-0050. The simulations were performed and analyzed at  the Leibniz Rechenzentrum Munich (LRZ) and at the
Barcelona Supercomputing Center (BSC). We thank DEISA for  giving  us  access
to  computing resources in these centers  through the DECI projects  SIMU-LU
and SIMUGAL-LU. We acknowledge the LEA Astro-PF collaboration and the ASTROSIM
network of the European Science Foundation (Science Meeting 2387) for the
financial support of the workshop ``The local universe: from dwarf galaxies to
galaxy clusters'' held in Jablonna near Warsaw in June/July 2009, where part of
this work was done. We also acknowledge MULTIDARK project (CSD2009-00064) and ASTROMADRID (S2009/ESP-146) for their support.

\bibliography{./ref}
\end{document}